\documentclass[conference]{IEEEtran}

\ifCLASSINFOpdf
\else
\fi

\usepackage[margin=1.25in]{geometry}
\usepackage{amsmath}
\usepackage{amsthm}
\usepackage{amssymb}
\usepackage{setspace}
\usepackage{color}
\usepackage{float}
\usepackage{dsfont}
\usepackage{enumerate}
\usepackage{graphicx}
\usepackage{sidecap}
\usepackage{subfigure}
\usepackage{algorithm}
\usepackage{algorithmic}
\newcommand{\commentout}[1]{}
\usepackage[utf8]{inputenc}
\usepackage{tikz}
\usetikzlibrary{plotmarks}
\usepackage{wrapfig}
\usepackage[normalem]{ulem}

\newtheorem{proposition}{Proposition}

\newtheorem{theorem}{Theorem}

\newtheorem{lemma}{Lemma}

\theoremstyle{definition}
\newtheorem{assumption}{Assumption}
\newtheorem{definition}{Definition}
\newtheorem{remark}{Remark}

\numberwithin{equation}{section}

\renewcommand{\subset}{\subseteq}
\renewcommand{\hat}{\widehat}
\renewcommand{\tilde}{\widetilde}
\renewcommand{\epsilon}{\varepsilon}

\def\SRF{\text{SRF}}

\newcommand{\calH}{\mathcal{H}}

\newcommand{\calR}{\mathcal{R}}
\newcommand{\calJ}{\mathcal{J}}

\newcommand{\diag}{\text{diag}}

\newcommand{\beq}{\begin{equation}}
\newcommand{\eeq}{\end{equation}}

\def\({\Big(}
\def\){\Big)}
\def\C{\mathbb{C}}

\def\T{\mathbb{T}}

\begin{document}
\title{Conditioning of restricted Fourier matrices and super-resolution of MUSIC }

\author{\IEEEauthorblockN{Weilin Li}
\IEEEauthorblockA{Courant Institute of Mathematical Sciences \\
New York University\\
Email: weilinli@cims.nyu.edu }
\and
\IEEEauthorblockN{Wenjing Liao	}
\IEEEauthorblockA{School of Mathematics\\
Georgia Institute of Technology\\
Email: wliao60@gatech.edu}
}

\maketitle

\begin{abstract}
This paper studies stable recovery of a collection of point sources from its noisy $M+1$ low-frequency Fourier coefficients. We focus on the super-resolution regime where the minimum separation of the point sources is below $1/M$. We propose a separated clumps model where point sources are clustered in far apart sets, and prove an accurate lower bound of the Fourier matrix with nodes restricted to the source locations. This estimate gives rise to a theoretical analysis on the super-resolution limit of the MUSIC algorithm.
\end{abstract}


\IEEEpeerreviewmaketitle

\section{Introduction}
%This paper studies stable recovery %of the locations and amplitudes 
%of $S$ point sources from its noisy low-frequency Fourier coefficients. 
In imaging and signal processing, $S$ point sources are usually represented by a discrete measure:
$
\mu(\omega)=\sum_{j=1}^S x_j\delta_{\omega_j}(\omega),
$
where
$x=\{x_j\}_{j=1}^S \in \mathbb{C}^S$ represents the source amplitudes and $\Omega=\{\omega_j\}_{j=1}^S\subset \mathbb{T} := [0,1)$ represents the source locations.  
A uniform array of $M+1$ sensors collects the noisy Fourier coefficients of $\mu$, denoted by $y \in \mathbb{C}^{M+1}$. %where $y_k$ is the noisy Fourier coefficient collected by the $k$-th sensor. 
One can write
\begin{equation}
	\label{eq:model3}
	y  = \Phi_M x+\eta,
\end{equation}
where $\Phi_M=\Phi_M(\Omega)$ is the $(M+1)\times S$ {\it Fourier} or {\it Vandermonde} matrix (with nodes on the unit circle):
\[
\Phi_M(\Omega)
=\begin{bmatrix}
1  & \ldots & 1 \\
e^{-2\pi i\omega_1} & \ldots & e^{-2\pi i\omega_S}\\
\vdots  & \vdots & \vdots \\
e^{-2\pi iM \omega_1}  & \dots & e^{-2\pi iM\omega_S} 
\end{bmatrix} ,
\]
and $\eta\in\C^{M+1}$ represents noise.

Our goal is to accurately recover $\mu$, especially the support $\Omega$, from $y$. 
%This inverse problem arises in many applications in imaging and signal processing, including inverse scattering, remote sensing, direction-of-arrival estimation and speech analysis.
%The key step is to estimate the support set $\Omega$. 
The measurements $y$ contains information about $\mu$ at a coarse resolution of approximately $1/M$, whereas we would like to estimate $\mu$ with a higher resolution. 
In the noiseless setting where $\eta=0$, the measure $\mu$ can be exactly recovered by many methods.
%the classical Prony's method \cite{prony1795essai} can recover $\mu$ exactly. However, Prony's method is unstable to noise. 
%
With noise, the stability of this inverse problem depends on $\Omega$. A crucial quantity is the {\it minimum separation} between the two closest points in $\Omega$, defined as
\[
\Delta
=\Delta(\Omega)
=\min_{1\leq j<k\leq S} |\omega_j-\omega_k|_\T,
\]
where $|\cdot|_\T$ is the metric on the torus $\T$. In imaging, $1/M$ %is typically called the {\it Rayleigh Length} ($RL$), and it 
is regarded as the standard resolution.
%minimum separation between two point sources that a standard imaging system can resolve \cite{den1997resolution}.
As a manifestation of the  Heisenberg uncertainty principle, recovery is sensitive to noise whenever $\Delta<1/M$, which case is referred as super-resolution.  %Super-resolution refers to the case where $\Delta < 1/M$. 
The {\it super-resolution factor} (SRF) is $M/\Delta$, standing for the maximum number of points in $\Omega$ that is contained in an interval of length $1/M$.

Prior mathematical work on super-resolution can be placed in three main categories: (a) the min-max error of super-resolution was studied in \cite{donoho1992superresolution,demanet2015recoverability} when point sources are on a fine grid of $\mathbb{R}$;
%addressed super-resolution from an information theoretic view. They considered the situation where the point sources are located on a grid on $\mathbb{R}$ with spacing $1/N$ and the given information consists of noisy continuous Fourier measurements. They both derived lower and upper bounds for a min-max error. These results are asymptotic as the grid spacing needs to be sufficiently small and the constants in the bounds are not explicitly given.
%
(b) when $\Omega$ is {\it well-separated} such that $\Delta\geq C/M$ for some constant $C>1$, some representative methods include total variation minimization (TV-min) \cite{candes2013super,duval2015exact, %fernandez2013support, azais2015spike, duval2015exact, 
li2017elementary}, greedy algorithms \cite{fannjiang2012coherence}, and subspace methods \cite{liao2016music,moitra2015matrixpencil}.
%hua1990matrixpencil,fannjiang2011music,liao2016music,liao2015multi,moitra2015matrixpencil,kailath1989esprit,schmidt1986multiple
These results address the issue of discretization error \cite{%fannjiang2011spie, 
fannjiang2012coherence} arising in sparse recovery, %\cite{candes2006robust,donoho2006compressed}, 
but they do not always succeed when $\Delta<1/M$; % \textcolor{red}{This sounds very critical. Perhaps we should say that these algorithms do not always succeed when $\Delta<1/M$.}; %if they cannot deal with the case $\Delta < 1/M$; 
(c) when $\Delta<1/M$, certain assumptions on the signs of $\mu$ are required by many optimization-based methods \cite{morgenshtern2016super,denoyelle2017support,benedetto2018super}. 
%requires that the sign of $\mu$ equals a dual polynomial on the support of $\mu$. 
Alternatively, {\it subspace methods} exploit a low-rank factorization of the data and can recover {\it complex} measures, but there are many unanswered questions related to its stability that we would like to address. 

This paper focuses on a highly celebrated subspace method, called MUltiple SIgnal Classification (MUSIC) \cite{schmidt1986multiple}. 
% Prior works \cite{donoho1992superresolution,demanet2015recoverability} addressed super-resolution from an information theoretic view. They considered the situation where the point sources are located on a grid on $\mathbb{R}$ with spacing $1/N$ and the given information consists of noisy continuous Fourier measurements. They both derived lower and upper bounds for a min-max error. These results are asymptotic as the grid spacing needs to be sufficiently small and the constants in the bounds are not explicitly given.
An important open problem is to understand the {\it super-resolution limit} of MUSIC: characterize the support sets  $\Omega$ and noise level for which MUSIC can stably recover all measures $\mu$ supported in $\Omega$ within a prescribed accuracy. 
Prior numerical experiments in \cite{liao2016music} showed that MUSIC can succeed even when $\Delta<1/M$, but a rigorous justification was not provided. This is one of our main motivations for the theory presented in this paper and in our more detailed preprint \cite{li2017stable}. 

As a result of Wedin's theorem \cite{wedin1972perturbation,li1998relative}, the stability of MUSIC obeys, in an informal manner,
	\begin{equation*}
	%\label{eqinformal}
	\text{Sensitivity}
	\leq \underbrace{%{\rm Constant} %\ \cdot \left(\frac{x_{\max}\sigma_{\max}^2(\Phi)}{x_{\min}\sigma_{\min}^2(\Phi)} \right)^k
		\frac{{\rm Constant}}{x_{\min}\sigma_{\min}^2(\Phi_{M})}}_{\text{ Noise amplification factor}} \cdot \underbrace{Q(\eta)}_{\text{Noise term}}, 
	\end{equation*}
	where $x_{\min} = \min_{j} |x_j|$, $\sigma_{\min}(\Phi_{M})$ is the smallest non-zero singular value of $\Phi_{M}$, and $Q(\eta)$ is a quantity depending on noise. Therefore, MUSIC can accurately estimate $\mu$ provided that the noise term is sufficiently small compared to the noise amplification factor which depends crucially on $\sigma_{\min}(\Phi_{M})$.

%Having highlighted the key role that $\sigma_{\min}(\Phi)$ plays in analyzing the resolution limit of MUSIC, we need to obtain an accurate lower bound for $\sigma_{\min}(\Phi)$.
In the separated case $\Delta>1/M$, accurate estimates for $\sigma_{\min}(\Phi_{M})$ and $\sigma_{\max}(\Phi_M)$ are known \cite{montgomery1974hilbert,vaaler1985some,moitra2015matrixpencil,liao2016music}. 	
In the super-resolution regime $\Delta<1/M$, the value of $\sigma_{\min}(\Phi_M)$ is extremely sensitive to the ``geometry" or configuration of $\Omega$, and a more sophisticated description of the ``geometry" of $\Omega$ other than the minimum separation is required. Based on this observation, we define a {\it separated clumps} model where $\Omega$ consists of well-separated subsets, where each subset contains several closely spaced points. This situation occurs naturally in applications where point sources clustered in far apart sets. 

Under this separated clumps model, we provide a lower bound of $\sigma_{\min}(\Phi_M)$ with the dominant term scaling like $\SRF^{-\lambda+1}$, where $\lambda$ is the cardinality of the largest clump. This is a significant improvement on existing lower bounds with continuous measurements where the exponents depend on the total sparsity $S$ \cite{donoho1992superresolution,demanet2015recoverability}. We use this estimate to rigorously establish the resolution limit of MUSIC and explain numerical results. More comprehensive explanations, comparisons, simulations, and proofs can be found in \cite{li2017stable}.

\section{Minimum singular value of Vandermonde matrices}

We first define a geometric model of $\Omega$ where the point sources are clustered into far apart clumps.

\begin{assumption}[Separated clumps model]
	\label{def:clumps}
	Let $M$ and $A$ be a positive integers and $\Omega\subset\T$ have cardinality $S$. We say that $\Omega$ consists of $A$ {\it separated clumps} with parameters $(M,S,\alpha,\beta)$ if the following hold.
	\begin{enumerate}
		\item 
		$\Omega$ can be written as the union of $A$ disjoint sets $\{\Lambda_a\}_{a=1}^A$, where each {\it clump} $\Lambda_a$ is contained in an interval of length $1/M$. 
		\item 
		$\Delta\geq\alpha/M$ with $\max_{1\leq a\leq A} (\lambda_a-1) < {1}/{\alpha}$ where $\lambda_a$ is the cardinality of $\Lambda_a$.
		\item 
		If $A>1$, then the distance between any two clumps is at least $\beta/M$.
	\end{enumerate}
\end{assumption}

\commentout{
\begin{definition}[separated clumps]
	\label{def:clumps}
	A set $\Omega\subset\T$ consists of {\it separated clumps} if there is an integer $A$ such that $\Omega$ can be written as the union of disjoint sets, denoted $\{\Lambda_a\}_{a=1}^A\subset\T$, where each {\it clump} $\Lambda_a$ is contained in an open interval of length $1/M$, and the distance between any two clumps exceeds $1/M$. Let $\lambda_a$ denote the cardinality of $\Lambda_a$. 
\end{definition}
}

\commentout{
To accurately estimate $\sigma_{\min}(\Phi)$ under the separated clumps model, we must consider both the intra-clump and inter-clump distances. We found that the following notion quantifies the local geometry of $\Omega$. 

\begin{definition}[Complexity]
	\label{def:complexity}
	The {\it complexity} at $\omega_j\in\Omega$ is the quantity,
	\[
	\rho_j
	%=\rho_j(\Omega,M)
	=\prod_{\omega_k\in\Omega\colon 0<|\omega_k-\omega_j|_\T<1/M} \frac{1}{\pi M|\omega_j-\omega_k|_\T}. 
	\]
\end{definition}}

There are many types of discrete sets that consist of separated clumps. Extreme examples include when $\Omega$ is a single clump containing all $S$ points, and when $\Omega$ consists of $S$ clumps containing a single point. While our theory applies to both extremes, the in-between case where $\Omega$ consists of several clumps each of modest size is the most interesting, and developing a theory of super-resolution for this case has turned out to be quite challenging.

Under this separated clumps model, we expect $\sigma_{\min}(\Phi_M)$ to be an $\ell^2$ aggregate of $A$ terms, where each term only depends on the ``geometry" of each clump. 

\commentout{

\begin{theorem}
	\label{thm:clump1}
	Let $M\geq 2S^2$. Assume $\Omega$ consists of separated clumps (Definition \ref{def:clumps}). If there are multiple clumps, assume that the distance between any two is at least
	\begin{equation*}
	%\label{eq:sep1}
	\max_{1\leq a\leq A} \max_{\omega_j\in\Lambda_a} \frac{10 \lambda_a^{5/2} (S\rho_j)^{1/(2\lambda_a)}}{M}.
	\end{equation*}
	%For each $1\leq a\leq A$, 
	Then there exist constants 
	$B_a := B_a(\lambda_a,M)$ for $a=1,\ldots,A$, such that
	%\[
	%B_a =B_a(\lambda_a,M)
	%=\frac{20\sqrt 2}{19} \(1-\frac{\pi^2}{3\lambda_a^2}\)^{-(\lambda_a-1)/2} \(\frac{M}{\lambda_a}\)^{\lambda_a-1} \Big\lfloor \frac{M}{\lambda_a}\Big\rfloor^{-(\lambda_a-1)}.\]
	%Let $\Phi=\Phi(\Omega,M)$ be the $(M+1)\times S$ Vandermonde matrix associated with $\Omega$. Then
	\begin{equation*}
	%\label{eq:clump1}
	\sigma_{\min}(\Phi_M(\Omega))
	\geq \sqrt{M} \(\sum_{a=1}^A \sum_{\omega_j\in \Lambda_a} (B_a \lambda_a^{\lambda_a-1} \rho_j)^2 \)^{-1/2}. 
	\end{equation*}
\end{theorem}

%\begin{remark}
%	\label{rem:Ba}
	The concrete form of $B_a$ is given in \cite{li2017stable}. %It only depends on $\lambda_a$ and $M$.% is therefore insensitive to the geometry of each $\Lambda_a$. 
	We can think of $B_a$ as a small universal constant. %because rounding becomes increasingly negligible as $M/\lambda_a$ increases, and the function $n\mapsto (1-\pi^2/(3n^2))^{-(n-1)/2}$ defined on the integers $n\geq 2$ approaches a horizontal asymptote of $1$ very quickly as $n$ increases.
	When each $\lambda_a$ is of moderate size and $M/\lambda_a$ is large, $B_a \approx 20\sqrt 2/19\approx 1.4886$. \commentout{(see Figure \ref{FigFunCa})}
%\end{remark}

\begin{remark}
	Although this is not the main point of Theorem \ref{thm:clump1}, we can apply it to the well-separated case. Assume that $\Delta\geq 10S^{1/2}/M$. Then each clump $\Lambda_a$ contains a single point, $B_a=20\sqrt 2/19$ for each $1\leq a\leq A$, and $\rho_j=1$ for each $\omega_j\in\Omega$. We readily check that the conditions of the theorem are satisfied, and thus,
	\[
	\sigma_{\min}(\Phi)
	\geq \frac{19}{20\sqrt 2} \sqrt{M}. 
	\]
	This shows that $\sigma_{\min}(\Phi)$ is on the order of $\sqrt M$ if $\Delta$ is about $\sqrt{S}$ times larger than $1/M$. This result is weaker than the one obtained in \cite{moitra2015matrixpencil}, which was derived using tools that specialized to the well-separated case. Note that $\sqrt M$ is approximately the largest $\sigma_{\min}(\Phi)$ can be because $\sigma_{\max}(\Phi)\leq \|\Phi\|_F = \sqrt{MS}$, where $\|\cdot\|_F$ is the Frobenius norm. 
\end{remark}

%Theorem \ref{thm:clump1} provides us with a lower bound for $\sigma_{\min}(\Phi)$ in terms of the complexities of $\Omega$. One might wonder what the bound reduces to in a more concrete situation. Suppose that $\Omega$ consists of $A$ separated clumps, but 
Additionally, if we assume $\Delta \ge \alpha/M$ for some $0<\alpha<1$ (note that $\SRF=1/\alpha$), we can upper bound the complexities of each $\omega_j$. The lower bound of $\sigma_{\min}(\Phi)$ becomes more concrete.
} 

%and determine the sufficient inter-clump separation. 

\begin{theorem}
	\label{thm:clump2}
	Let $M\geq S^2$. Assume $\Omega$ satisfies Assumption \ref{def:clumps} with parameters $(M,S,\alpha,\beta)$ for some $\alpha>0$ and 
	\begin{equation}
	\label{eq:sep2}
	\beta\geq \max_{1\leq a\leq A} \frac{20S^{1/2}\lambda_a^{5/2}}{\alpha^{1/2}}.
	\end{equation}
	\commentout{
	 has cardinality $S$, consists of $A$ separated clumps, and $\Delta\geq\alpha/M$ with $\max_{1\leq a\leq A} (\lambda_a-1) < {1}/{\alpha}$. If $\Omega$ consists of multiple clumps such that $A>1$, assume that the distance between any two is at least
	\begin{equation}
	\label{eq:sep2}
	\max_{1\leq a\leq A}\  \frac{20S^{1/2}\lambda_a^{5/2}}{\alpha^{1/2}M}.
	\end{equation}}
	%For each $1\leq a\leq A$, let $B_a=B_a(\lambda_a,M)$ be the constant defined in Theorem \ref{thm:clump1} and
	%\begin{equation}
	%\label{thm2ca}
	%C_a=C_a(\lambda_a,M)= B_a\ \(\frac{\lambda_a}{\pi}\)^{\lambda_a-1}\(\sum_{j=1}^{\lambda_a} \prod_{k=1,\ k\not=j}^{\lambda_a} \frac{1}{(j-k)^2}\)^{1/2}.
	%\end{equation}
	Then there exist explicit constants $C_a>0$ such that
	\begin{equation}
	%\label{thm2lowerbound}
	\hspace{-.5em} \sigma_{\min}(\Phi_M)
	\geq \sqrt{M}\(\sum_{a=1}^A \big( C_a \alpha^{-\lambda_a+1} \big)^2 \)^{-\frac{1}{2}}. 
	\end{equation}
\end{theorem}

%\begin{remark}
%	\label{remark:thm12}
	%We would like to compare the assumptions and statements of Theorems \ref{thm:clump1} and \ref{thm:clump2}. The latter is more concrete since it bounds $\sigma_{\min}(\Phi)$ in terms of $\alpha=1/\SRF$, but it is less accurate. Suppose each clump $\Lambda_a$ consists of $\lambda_a$ points equispaced by $\alpha/M$. Then the lower bounds \eqref{eq:clump1} and \eqref{thm2lowerbound} are identical. For all other configurations of $\Omega$, the estimate \eqref{eq:clump1} is more accurate than \eqref{thm2lowerbound}. The separation condition \eqref{eq:sep1} is always weaker than \eqref{eq:sep2}. As mentioned in Remark \ref{rem:Ba}, the constant $B_a$ weakly depends on $\lambda_a$. The constant $C_a$ scales like $\lambda_a^{\lambda_a-1}$ for large $\lambda_a$. %This discrepancy arises because $\rho_j$ is significantly different $\alpha^{-\lambda_a+1}$ when $\lambda_a$ is large. 
%\end{remark}

The main feature of this theorem is the exponent on $\SRF=1/\alpha$, which depends on the cardinality of each clump as opposed to the total number of points. Let $\lambda$ be the cardinality of the largest clump: $\lambda = \max_{a=1}^A \lambda_a$.
%As an example, let us consider the special case where each clump contains $\lambda$ points consecutively spaced by $\alpha/M$ and %the distance between clumps is $\beta/M$ where $\beta$ is large enough so that \eqref{eq:sep2} holds 
%the inter-clump distance is large enough so that condition \ref{eq:sep2} is satisfied.
%(see Figure \ref{FigDemoClumps1}).
\commentout{
\begin{figure}[h]
	\centering
	\begin{tikzpicture}[xscale = 0.6,yscale = 0.5]
	\draw[thick] (-6,0) -- (-0.5,0);
	\filldraw[red] (-5,0) circle (0.1cm);		
	\filldraw[red] (-4.7,0) circle (0.1cm);		
	\filldraw[red] (-4.4,0) circle (0.1cm);		
	\draw[blue,thick,<->] (-4.7,-0.2) -- (-4.4,-0.2);
	\node[blue,below] at (-4.4,-0.2) {$\alpha/M$};
	
	%\draw[blue,thick,<->] (-4.4,0.2) -- (-2,0.2);
	%\node[blue,above] at (-3.2,0.2) {$\beta/M$};
	
	\node[above] at (-4.7,0.2) {$\Lambda_1$};
	\filldraw[red] (-2,0) circle (0.1cm);		
	\filldraw[red] (-1.7,0) circle (0.1cm);		
	\filldraw[red] (-1.4,0) circle (0.1cm);		
	\draw[blue,thick,<->] (-1.7,-0.2) -- (-1.4,-0.2);
	\node[blue,below] at (-1.4,-0.2) {$\alpha/M$};
	\node[above] at (-1.7,0.2) {$\Lambda_2$};
	\draw[dotted,thick] (-0.5,0) -- (1,0);
	\draw[thick] (1,0) -- (6.4,0);
	\filldraw[red] (2,0) circle (0.1cm);		
	\filldraw[red] (2.3,0) circle (0.1cm);		
	\filldraw[red] (2.6,0) circle (0.1cm);		
	\draw[blue,thick,<->] (2.3,-0.2) -- (2.6,-0.2);
	\node[blue,below] at (2.6,-0.2) {$\alpha/M$};
	\node[above] at (2.3,0.2) {$\Lambda_{A-1}$};
	\filldraw[red] (5,0) circle (0.1cm);		
	\filldraw[red] (5.3,0) circle (0.1cm);		
	\filldraw[red] (5.6,0) circle (0.1cm);		
	\draw[blue,thick,<->] (5.3,-0.2) -- (5.6,-0.2);
	\node[blue,below] at (5.6,-0.2) {$\alpha/M$};
	\node[above] at (5.3,0.2) {$\Lambda_A$};
	%\draw[blue,thick,<->] (2.6,0.2) -- (5,0.2);
	%\node[blue,above] at (3.8,0.2) {$\beta/M$};
	%%  
	%\node[below] at (-0.5,-1) {$\lambda =|\Lambda_1| = |\Lambda_2|= \ldots = |\Lambda_A| $};
	\end{tikzpicture}
	\caption{$\Omega = \cup_a \Lambda_a$ where each $\Lambda_a$ contains $\lambda$ points ($\lambda=3$ above) spaced consecutively by $\alpha/M$. %The distance between clumps is $\beta/M$. 
	}
	\label{FigDemoClumps1}
\end{figure}
}
 %The ratio between Rayleigh length and the minimum separation is the standard super-resolution factor (SRF). Here ${\rm SRF} = 1/\alpha$. 
Theorem \ref{thm:clump2} implies
\begin{equation}
\label{eqlowert1}
\sigma_{\min}(\Phi_M) \ge
C \sqrt{M}\ \SRF^{-\lambda+1}.
\end{equation}
Previous results \cite{donoho1992superresolution,demanet2015recoverability} strongly suggest (we avoid using ``imply" because they studied a similar inverse problem but with continuous, rather than discrete measurements like the ones considered here) that
\begin{equation}
\label{eqlowert2}
\sigma_{\min}(\Phi_M)\geq C\sqrt{M}\ \SRF^{-S+1}. 
\end{equation}
By comparing the inequalities \eqref{eqlowert1} and \eqref{eqlowert2}, we see that our lower bound is dramatically better when all of the point sources are not located within a single clump. These results are also consistent with our intuition that $\sigma_{\min}(\Phi_M)$ is smallest when $\Omega$ consists of $S$ closely spaced points; more details about this can be found in \cite{li2017stable}. {In \cite{batenkov2018stability}, a lower bound of $\sigma_{\min}(\Phi_M)$ is derived for a model called clustered nodes; a detail comparison between Theorem \ref{thm:clump2} and results in \cite{batenkov2018stability} can be found in \cite{li2017stable}.}

\commentout{
	=\(1-\frac{\pi^2}{3\lambda^2}\)^{-(\lambda-1)/2} \(\sum_{j=1}^{\lambda} \prod_{k=1,\ k\not=j}^{\lambda} {(j-k)^{-2}}\)^{1/2} \(\frac{M}{\lambda}\)^{\lambda-1} \Big\lfloor \frac{M}{\lambda}\Big\rfloor^{-(\lambda-1)}.

\begin{remark}
	\label{remark:batenkov} 
	In the process of revising the first draft of this manuscript, the authors of \cite{batenkov2018stability}, independent of our work, used different techniques to derive lower bounds for $\sigma_{\min}(\Phi)$ when $\Omega$ consists of clumps, see \cite[Definition 1.1]{batenkov2018stability} for their model. We point out the differences between their \cite[Corollary 1.1]{batenkov2018stability} and our Theorem \ref{thm:clump2}. 
	\begin{enumerate}[(a)]
		\item 
		They assume that $\Omega$ consists of clumps that are all contained in an interval of length approximately $1/S^2$. For ours, the clumps can be spread throughout $\T$ and not have to be concentrated on a sub-interval. 
		\item 
		They require the aspect ratio $M/S$ of the Vandermonde matrix $\Phi$ to be at least $4S^2$, whereas we only need at least $2S$. They also require an upper bound on $M/S$, which prohibits their Vandermonde matrix from being too tall. 
		\item 
		If $\lambda$ is the cardinality of the largest clump in $\Omega$, then they obtained the inequality $\sigma_{\min}(\Phi)\geq C_S\sqrt{M} \cdot \SRF^{-\lambda+1}$ for some $C_S$ depending only on $S$ which scales like $S^{-2S}$. Our implicit constant is more complicated, but it scales like $A^{-1/2}\lambda^{-\lambda}$. 
	\end{enumerate}
\end{remark}
}
%% upper bound

The following theorem provides an upper bound on $\sigma_{\min}(\Phi_M)$ when $\Omega$ contains $\lambda$ consecutive points spaced by $\alpha/M$, and this shows that the dependence on $\SRF$ in inequality \eqref{eqlowert1} is optimal.

%further shows that, if $\Omega$ contains $\lambda$ consecutive points with spacing $\alpha/M$ for a sufficiently small $\alpha$, there exists $\tilde{C}(\lambda,M)$ such that $\sigma(\Phi_M(\Omega)) \le \tilde{C}(\lambda,M)\sqrt{M}\alpha^{\lambda-1}$, which matches the lower bound in \eqref{eqlowert1}.

\begin{theorem}
	\label{prop:upper}
	Let $\lambda\leq S\leq M-1$, and there exists a constant $c>0$ depending only on $\lambda$ such that the following hold: for any $0<
	\alpha\leq {c ({M+1})^{-1/2}}$, $\omega\in\T$ and $\Omega\subset \T$ of cardinality $S$ that contains the set $\omega+\{0,{\alpha}/{M},\dots,{(\lambda-1)\alpha}/{M} \}$, we have $\sigma_{\min}(\Phi_M)
	\leq C_\lambda \alpha^{\lambda-1}$.
\end{theorem}

% % % % % % % % % % % % % % % % % % % %
\section{MUSIC and its super-resolution limit}
\label{sec:music}

\commentout{
	\begin{figure}[h]
		\centering
		\begin{tikzpicture}[xscale = 1,yscale = 1]
		\draw[thick] (-6,0) -- (5,0);
		\draw[red,thick,->] (-5,0) -- (-5,2);
		\draw[red,thick,->] (-4.7,0) -- (-4.7,2);
		\draw[red,thick,->] (-4.4,0) -- (-4.4,2);
		\draw[blue,thick,<->] (-4.7,-0.2) -- (-4.4,-0.2);
		\node[blue,below] at (-4.1,-0.2) {$\ge \alpha/M$};
		\draw[blue,thick,<->] (-4.4,1) -- (-1,1);
		\node[blue,above] at (-2.5,1) {$\ge \beta/M$};
		\draw[red,thick,->] (-1,0) -- (-1,2);
		\draw[red,thick,->] (-0.7,0) -- (-0.7,2);
		\draw[blue,thick,<->] (-1,-0.2) -- (-0.7,-0.2);
		\node[blue,below] at (-0.4,-0.2) {$\ge \alpha/M$};
		\draw[blue,thick,<->] (-0.7,1) -- (3,1);
		\node[blue,above] at (1.2,1) {$\ge \beta/M$};
		\draw[red,thick,->] (3,0) -- (3,2);
		\draw[red,thick,->] (3.4,0) -- (3.4,2);
		\draw[red,thick,->] (3.7,0) -- (3.7,2);
		\draw[blue,thick,<->] (3.4,-0.2) -- (3.7,-0.2);
		\node[blue,below] at (4,-0.2) {$\ge \alpha/M$};
		\node[above] at (-4.6,2.2) {$\Omega_1$};
		\node[above] at (-0.75,2.2) {$\Omega_2$};
		\node[above] at (3.4,2.2) {$\Omega_3$};
		\node[below] at (-0.5,-1) {$\gamma = \max(|\Omega_1|,|\Omega_2|,|\Omega_3|) = 3$};
		\end{tikzpicture}
		\caption{A demo of our separated clumps model}
		\label{FigDemoClumps}
	\end{figure}
}

%Many interests in imaging center on inventing super-resolution algorithms and understanding the resolution limit of these algorithms. 
In signal processing, the MUSIC algorithm \cite{schmidt1986multiple}, has been widely used due to its superior numerical performance among subspace methods. %It was well known that MUSIC has super-resolution phenomenon, i.e. the capability of resolving point sources separated below RL. The resolution limit of MUSIC was discovered by numerical experiments in \cite{liao2016music}, but has never been rigorously proved. By resolution limit we mean the relation between the geometry of the support set $\Omega$ and the noise level for which the recovery of all point sources is possible. 
%A main contribution of this paper is to prove the resolution limit of MUSIC.
MUSIC relies upon the Vandermonde decomposition of a Hankel matrix, and its stability to noise can be formulated as a matrix perturbation problem. 

Throughout the following exposition, we assume that $L$ is an integer satisfying the inequalities $S \le L \le M+1-S$. The Hankel matrix of $y$ is 
\[
\calH(y) = 
\begin{bmatrix}
y_0 & y_1 & \hdots & y_{M-L}
%\\
%y_1 & y_2 & \hdots & y_{M-L+1}
\\
\vdots & \vdots & \ddots & \vdots
\\
y_{L} & y_{L+1} & \hdots & y_{M}
\end{bmatrix}.% \in \mathbb{C}^{(L+1) \times (M-L+1)}.
\]
If we denote the noiseless measurement vector by $y^0 = \Phi_{M}(\Omega) x$, then it is straightforward to verify that we have the following Vandermonde decomposition
$$
\calH(y^0)  
= \Phi_L \diag(x_1,\dots,x_S)\Phi_{M-L}^T. 
$$
Observe that both $\Phi_L$ and $\Phi_{M-L}$ have full column rank when $S \le L \le M+1-S$ and that $H(y^0)$ has rank $S$. The Singular Value Decomposition (SVD) of $\calH(y^0)$ is of the form
\begin{equation*}
\calH(y^0) = [U \ W] \ 
{\text{diag}(\sigma_1,\ldots,\sigma_S,0,\ldots,0)}%_{L \times (M-L+1)} 
%\ [\underbrace{V_1}_{(M-L+1) \times S} \ \underbrace{V_2}_{(M-L+1) \times (M-L+1-S)}]^*
V^*,
%\label{svdnoiseless}
\end{equation*}
where %$U \in \C^{(L+1) \times S}$, $W \in \C^{(L+1) \times (L-S)}$, and 
$\sigma_1  \ge \ldots \ge \sigma_S$ are the non-zero singular values of $\calH(y^0)$.  The columns of $U \in \C^{(L+1)\times S}$ and $W \in \C^{(L+1)\times (L+1-S)}$ span $\text{Range}(\calH(y^0))$ and $\text{Range}(\calH(y^0))^\perp$ respectively, which are called the signal space and the noise space. 

%Note that $\Phi_L$ and $\Phi_{M-L}$ have full column rank and $\calH(y^0)$ has rank $S$. 
%

\renewcommand{\algorithmicrequire}{\textbf{Input:}}
\renewcommand{\algorithmicensure}{\textbf{Output:}}
\begin{algorithm}[t!]                      	% enter the algorithm environment
	\caption{MUltiple SIgnal Classification}          	% give the algorithm a caption
	\label{algorithmmusic}		% and a label for \ref{} commands later in the document
	\begin{algorithmic}[1]                    	% enter the algorithmic environment
		\vspace{0.25em}
		\REQUIRE  $y \in \C^{M+1}$, sparsity $S$, $L$.
		\STATE Form Hankel matrix $\calH(y) \in \C^{(L+1)\times (M-L+1)}$ \vspace{-1em}
		\STATE Compute the SVD of $\calH(y)$: 
		\begin{equation*}
		%\label{svdnoisy}
		\calH(y) = [{\hat U} \  \hat W] {{\rm diag}(\hat \sigma_1 , \ldots , \hat \sigma_S, \hat \sigma_{S+1} ,\ldots)}
		{\hat{V}}^*, \end{equation*}
		%where $\hat \sigma_1 \ge \hat \sigma_2  \ldots \ge \ldots$ are the singular values of $\calH(y)$.
		where ${\hat U} \in \C^{(L+1)\times S}$, $ \hat W \in \C^{(L+1)\times (L+1-S)}$.% \vspace{-0.1em}
		\STATE Compute the imaging function $\hat\calJ(\omega) = {\|\phi_L(\omega)\|_2}/{ \|{\hat W}^* \phi_L (\omega)\|_2}, \ \omega \in [0,1)$. \vspace{0.5em}
		\ENSURE $\hat\Omega= \{\hat\omega_j\}_{j=1}^S$ corresponding to the $S$ largest local maxima of  $\hat\calJ$.
	\end{algorithmic}
\end{algorithm}

%The MUSIC algorithm was proposed by Schmidt \cite{schmidt1986multiple}. %It amounts to finding the noise space of the Hankel matrix, forming a noise-space correlation function (or its reciprocal which is called the imaging function), and identifying the $S$ smallest local minima of the noise-space correlation function (or the $S$ peaks of the imaging function) as the support set.

For any $\omega\in\T$ and positive integer $L$, we define the {\it steering vector} of length $L+1$ to be 
\[
\phi_L(\omega) = [1 \ e^{-2\pi i \omega} \ e^{-2\pi i 2 \omega} \ \ldots \ e^{-2\pi i L \omega}]^T.
%\in \C^{L+1}. 
\]
%and then $\Phi_L = [\phi_L(\omega_1) \ \ldots \ \phi_L(\omega_S) ]$. 
%so that the range of $\calH(y^0)$ is exactly spanned by $\{\phi_L(\omega_1), \ldots , \phi_L(\omega_S)\}$. 
%
MUSIC is based on the following observation that
$$\omega \in \Omega
\text{ iff }
\phi_L(\omega) \in %\text{Range}(\Phi_L) = 
\text{Range}(\calH(y^0)) = \text{Range}(U).
$$

\vspace{-0.15cm}
\begin{table}[h!]
\centering
{ Table I: Functions in the MUSIC algorithm}
	\vspace{0.2cm}\\
	\renewcommand{\arraystretch}{1.5}
	\centering
	\resizebox{1\columnwidth}{!}{
		\begin{tabular}{ |c | c  |c| }
			\hline
			&   Noise-space correlation function&  Imaging function\\
			[5pt]
			\hline
			Noiseless 
			& $\calR(\omega) = \frac{\| W^* \phi_L(\omega)\|_2}{\|\phi_L(\omega) \|_2}$
			&   $\calJ(\omega) = \frac{1}{\calR(\omega)} %= \frac{\|\phi_L(\omega) \|_2}{\| W^* \phi_L(\omega)\|_2}
			$
			\\
			[5pt]
			\hline
			%   [15pt]
			% Measure & $\rho(\Cjk)$  & $\hrho(\Cjk) = \hnjk /n$
			Noisy    & 
			$\hat\calR(\omega) = \frac{\| {\hat W}^* \phi_L(\omega)\|_2}{\|\phi_L(\omega) \|_2}$
			& 
			$\hat\calJ(\omega) = \frac{1}{\hat\calR(\omega)} %= \frac{\|\phi_L(\omega) \|_2}{\| {\hat W}^* \phi_L(\omega)\|_2}
			$
			\\
			[2pt]
			\hline
		\end{tabular}
	}
	%\caption{}%Noise-space correlation and imaging functions}
	%\label{TableMUSICFunctions}
\end{table}

This observation can be reformulated in terms of the noise-space correlation function $\calR(\omega)$ and the imaging function $\calJ(\omega)$ (see Table I for their definitions), as summarized in the following lemma.

\begin{lemma}
	Let $S \le L \le M+1-S$. Then 
	\[
	\omega \in \{\omega_j\}_{j=1}^S
	\Longleftrightarrow
	\calR(\omega) = 0
	\Longleftrightarrow
	\calJ(\omega) = \infty.
	\]
\end{lemma}

%The following lemma is a basis for the MUSIC algorithm in the noiseless case. 
%
To summarize this discussion: in the noiseless case where we have access to $y^0$, the source locations can be exactly identified through the zeros of the noise-space correlation function $\calR(\omega)$
or the peaks of the imaging function $\calJ(\omega)$. 

In the presence of noise, we only have access to $\calH(y)$, which is a perturbation of $H(y^0)$:
$$\calH(y) = \calH(y^0) + \calH(\eta).$$
The noise-space correlation and imaging functions are perturbed to 
$\hat\calR(\omega)$ and
$\hat\calJ(\omega)$
respectively. Stability of MUSIC depends on the perturbation of the noise-space correlation function from $\calR(\omega)$ to $\hat{\calR}(\omega)$ which we measure by 
\[
\|\hat\calR - \calR\|_\infty := \max_{\omega \in [0,1)} |\hat\calR(\omega) - \calR(\omega)|.
\]
By using Wedin's theorem \cite[Theorem 3.4]{wedin1972perturbation,li1998relative}, we can prove the following perturbation bound. 
\begin{proposition}
	\label{lemmapermusic1}
	Let $S \le L \le M+1-S$. Suppose $2\|\calH(\eta)\|_2 < x_{\min}\sigma_{\min}(\Phi_L)\sigma_{\min}(\Phi_{M-L})$. Then
	$$	
	\|\hat\calR - \calR\|_\infty 
	\le \frac{2\|\calH(\eta)\|_2}{x_{\min}\sigma_{\min}(\Phi_L)\sigma_{\min}(\Phi_{M-L})}.	
	$$
\end{proposition}

%Figure \ref{Fig_MUSICThreeSpikes} shows that, the imaging function $\hat\calJ(\omega)$ still peaks around the true sources, as long as the noise-to-signal ratio is low enough. However, MUSIC can fail when the noise-to-signal ratio increases.

If $\eta$ is independent Gaussian noise, i.e., $\eta \sim \mathcal{N}(0,\sigma^2 I)$, the spectral norm of $\calH(\eta)$ satisfies the following concentration inequality \cite[Theorem 4]{liao2015multi}:
\begin{lemma}
	If $\eta \sim \mathcal{N}(0,\sigma^2 I)$, then
	\label{lemmanoise}
	\begin{align*}
	\mathbb{E} \|\calH(\eta)\|_2 
	&\le \sigma \sqrt{2 C(M,L)\log(M+2)},
	\\
	\mathbb{P}\left\{ 
	\|\calH(\eta)\|_2 \ge t
	\right\}
	&\le (M+2) \exp\left(-\frac{t^2}{2\sigma^2 C(M,L)}\right),
	\end{align*}  
	for $t>0$, and $C(M,L) =\max(L+1,M-L+1)$.
\end{lemma}

%The crucial quantities that affect the stability of MUSIC are $\sigma_{\min}(\Phi_L)$ and $\sigma_{\min}(\Phi_{M-L})$. 
%For the rest of this paper, we set $ L=\lfloor M/2\rfloor$ to balance $\sigma_{\min}(\Phi_L)$ and $\sigma_{\min}(\Phi_{M-L})$. Theorem \ref{thm:clump2} gives an accurate estimate for the minimum singular value of $\sigma_{\min}(\Phi_L)$ under the separated clumps model.
%
Combining Proposition \ref{lemmapermusic1}, Lemma \ref{lemmanoise} and Theorem \ref{thm:clump2} gives rise to a stability analysis of MUSIC:

\begin{theorem}
	\label{thmmusic}
	Let $M$ be an even integer satisfying $M \ge 2 S^2$ and set $L=M/2$. Fix parameters $\epsilon>0$, $\nu>1$, and let $\eta \sim \mathcal{N}(0,\sigma^2 I)$. 
	Assume $\Omega$ satisfies Assumption \ref{def:clumps} with parameters $(L,S,\alpha,\beta)$  for some $\alpha>0$ and $\beta$ satisfying \eqref{eq:sep2}.
%Assume $\Omega\subset\T$ has cardinality $S$, consists of $A$ separated clumps, and $\Delta\geq\alpha/M$ with $\max_{1\leq a\leq A} (\lambda_a-1) < {1}/{\alpha}$. If $A>1$, assume that the distance between any two clumps is at least twice the quantity given in \eqref{eq:sep2}. 
	There exist explicit constants $c_a>0$ such that if
	\begin{align*}
	\frac{\sigma}{x_{\min}} 
	&< C(M,\nu) \Big(\sum_{a=1}^A c_a^2  \alpha^{-2(\lambda_a-1)} \Big)^{-1} \epsilon, \\
	C(M,\nu) 
	&=\frac{M}{32 \sqrt{\nu(M+2)\log(M+2)}},
	%\label{thmmusiceq1}
	\end{align*}
	%Let $\hat{\calR}$ and $\calR$ be the noise-space correlation functions	in MUSIC with $L = M/2$ in the noisy and noiseless case, respectively. Then,
	then with probability no less than $1-(M+2)^{-(\nu-1)}$,
	$$	\|\hat\calR - \calR\|_\infty 
	\le \epsilon. 
	% \frac{3200\sqrt{\nu(M+2)\log(M+2)}}{361M} 
	% \sum_{a=1}^A C_a^2 \lambda_a^2 \left(\frac{2\lambda_a}{\pi \alpha} \right)^{2\lambda_a-2}\cdot \frac{\sigma}{x_{\min}},
	$$
\end{theorem}

In order to guarantee an $\epsilon$-perturbation of the noise-space correlation function, the noise-to-signal ratio should follow the scaling law
\begin{equation*}
%\label{musicres1}
\frac{\sigma}{x_{\min}} \propto \sqrt{\frac{M}{\log M}} \left(
\sum_{a=1}^A c^2_a  \alpha^{-2(\lambda_a-1)}
\right)^{-1}\epsilon.
\end{equation*}
Let $\lambda$ be the cardinality of the largest clump. By \eqref{eqlowert1}, this scaling law reduces to
\begin{equation*}
\frac{\sigma}{x_{\min}} \propto \sqrt{\frac{M}{\log M}} \alpha^{2\lambda-2}\epsilon
= \sqrt{\frac {M}{\log M}} \ {\rm SRF}^{-(2\lambda-2)}\epsilon. 
%\label{musicres2}
\end{equation*}
The resolution limit of MUSIC is exponential in ${\rm SRF}$, but the exponent only depends on the cardinality of the separated clumps instead of the total sparsity $S$. These estimates are verified by numerical experiments in \cite{li2017stable}.
%When noise is i.i.d. gaussian, increasing $M$ helps to improve the resolution limit of subspace methods.

\commentout{
\subsection{ESPRIT}
ESPRIT stands for Estimation of Signal Parameters via Rotational Invariance Techniques, which is a high-resolution algorithm proposed by Roy and Kailath \cite{kailath1989esprit}. It does  not  involve  an  exhaustive  search 
through all possible steering vectors to estimate the point sources and dramatically reduces the computational  and  storage  requirements  compared  to  MUSIC. The  goal  of  ESPRIT is  to  exploit the rotational  invariance in the signal subspace which is created by two arrays with a translational invariance structure. Here we will state the single-snapshot ESPRIT algorithm from \cite{fannjiang2016compressive}.

In the noiseless case, $\calH(y^0) = \Phi_L X \Phi_{M-L}^T$. Let $H_0$ and $H_1$ be the submatrices containing the first $L$ and the last $L$ rows of $\calH(y^0)$, respectively. Then
\begin{align*}
H_0 &= \Phi_{L-1} X \Phi_{M-1}^T
\\
H_1 &= \Phi_{L-1} D_\omega X \Phi_{M-1}^T
\end{align*}
where $D_\omega = \diag(\{e^{-2\pi i \omega_j}\}_{j=1}^S)$. One can verify that 
$$H_1 = \Phi_{L-1} D_\omega \Phi_{L-1}^\dagger H_0.$$
Let $\Psi =  \Phi_{L-1} D_\omega \Phi_{L-1}^\dagger$, whose $S$ nonzero eigenvalues are exactly $\{e^{-2\pi i \omega_j}\}_{j=1}^S$. The ESPRIT technique amounts to finding the support set through the eigenvalues of $\Psi$, which can be explicitly computed as $\Psi = H_1 H_0^\dagger$. 
Suppose the SVD of $H_0$ is given by 
\begin{equation*}
H_0 = \underbrace{U_0}_{L \times S} \ 
\underbrace{\Sigma_0}_{S \times S}\
\underbrace{V_0^*}_{S\times (M-L+1)}
\text{ where } \Sigma_0 =  \underbrace{\text{diag}(\sigma_1(H_0),\ldots,\sigma_S(H_0))}_{S \times S}
\end{equation*}
and then $H_0^\dagger = V_0 \Sigma_0^{-1} U_0^*$.

\begin{lemma}
	\label{lemmaesprit1}
	Suppose $M+1 \ge 2S$ and $L$ is chosen such that $L \ge S$ and $M-L+1\ge S$. Then the $S$ nonzero eigenvalues of $\Psi = H_1 H_0^\dagger$ are exactly $\{e^{-2\pi i \omega_j}\}_{j=1}^S$.
\end{lemma}

\renewcommand{\algorithmicrequire}{\textbf{Input:}}
\renewcommand{\algorithmicensure}{\textbf{Output:}}
\begin{algorithm}[t!]                      	% enter the algorithm environment
	\caption{ESPRIT}          	% give the algorithm a caption
	\label{algorithmesprit}		% and a label for \ref{} commands later in the document
	\begin{algorithmic}[1]                    	% enter the algorithmic environment
		\REQUIRE  $y \in \C^{M+1}$, sparsity $S$, $L$
		\STATE Form Hankel matrix $\calH(y) \in \C^{(L+1)\times (M-L+1)}$. Let $\hat H_0$ and $\hat H_1$ be the submatrices of $\calH(y)$ containing the first $L$ and the last $L$ rows respectively.
		\STATE Compute $\hat \Psi = \hat H_1 \hat H_0^\dagger$
		where $\hat H_0^\dagger = \hat V_0 \hat\Sigma_0^{-1} \hat U_0^*$ with each term given by
		the SVD of $\hat H_0$:
		\begin{align*}
		\hat H_0 =  [\underbrace{\hat U_0}_{L \times S} \  \underbrace{\hat W_0}_{L \times (L-S)}] \underbrace{{\rm diag}(\sigma_1(\hat H_0) , \ldots , \sigma_S(\hat H_0), \sigma_{S+1}(\hat H_0) ,\ldots)}_{L \times (M-L+1)} [\underbrace{\hat V_0}_{(M-L+1) \times S} \ \underbrace{\cdots}_{(M-L+1) \times (M-L+1-S)}]^*
		\end{align*}
		and $\hat\Sigma_0 = {\rm diag}(\sigma_1(\hat H_0) , \ldots , \sigma_S(\hat H_0)) \in \mathbb{C}^{S \times S}$.
		\STATE Compute the $S$ largest eigenvalues of $\hat\Psi$, denoted by $\hat f_1,\ldots,\hat f_S$.
		\ENSURE $\hat \Omega = \{\hat\omega_j\}_{j=1}^S$ where $\omega_j = -\frac{\angle \hat f_j}{2\pi }$.
	\end{algorithmic}
\end{algorithm}

In the noisy case, we state the ESPRIT algorithm in Algorithm \ref{algorithmesprit}. Reconstruction error of ESPRIT is measured by the bottleneck distance  $${\rm dist}_B(\Omega , \hat\Omega) 
:= \inf_{\substack{\text{bijection }  \psi:\  \Omega\rightarrow  \hat\Omega } }  \ \sup_{\hat \omega \in \Omega} \left |\hat \omega - \psi(\omega)\right|_{\mathbb{T}}.$$
The stability of ESPRIT is given by the following proposition, which first appears in \cite{fannjiang2016compressive}.

\begin{proposition}
	\label{propesprit}
	Suppose $M+1 \ge 2S$ and $L$ is chosen such that $L \ge S$ and $M-L+1\ge S$. Let $\hat\Omega = \{\hat\omega_j\}_{j=1}^S$ be the recovered support by ESPRIT. If $2\|\calH(\eta)\| \le x_{\min}\sigma_S(\Phi_{L-1})\sigma_S(\Phi_{M-L})$, then
	%${\rm dist}_B(\Omega , \hat\Omega) := \inf_{{\text{bijection }  \psi:\  \Omega\rightarrow  \hat\Omega } }  \sup_{\hat \omega \in \Omega} \left |\hat \omega - \psi(\omega)\right|$  
	\begin{equation}
	\label{propespriteq}
	{\rm dist}_B(\Omega , \hat\Omega) 
	\le
	\frac{10x_{\max}\sigma_{\max}(\Phi_{L-1})\sigma_{\max}(\Phi_{M-L})  }{ \pi x_{\min}\sigma_{\min}(\Phi_{L-1})\sigma_{\min}(\Phi_{M-L})}\cdot
	\frac{ \|\calH(\eta)\|_2}{ x_{\min}\sigma_{\min}^2(\Phi_{L-1})\sigma_{\min}^2(\Phi_{M-L})}.
	\end{equation}
\end{proposition}

\textcolor{red}{I think Theorem 3 in http://www.ams.jhu.edu/~fill/papers/MoorePenrose.pdf might be helpful. Let $H_0(\eta)$ be the first $L-1$ rows of the Hankel matrix of the noise. If the noise level is sufficiently small so that the rank of $H_0$ and $\tilde H_0$ are identical, then by the referenced theorem,
	\begin{align*}
	\tilde H_0^\dagger
	&=(H_0+H_0(\eta))^\dagger \\
	&=(I-S) H_0^\dagger (I-T)-S(H_0(\eta))^\dagger T \\
	&=H_0^\dagger - SH_0^\dagger -H_0^\dagger T + S(H_0^\dagger-H_0(\eta)) T,
	\end{align*}
	where 
	\[
	S=(P_{R(H_0(\eta)^*)} P_{R(H_0^*)^\perp})^\dagger
	\quad \text{and}\quad 
	T=(P_{R(H_0)^\perp} P_{R(H_0(\eta))})^\dagger.
	\]
	This already looks similar to how Wedin's theorem is derived. Using this equation, we have
	\begin{align*}
	\tilde\Psi -\Psi
	&=\tilde H_1\tilde H_0^\dagger - H_1H_0^\dagger \\
	&=(\tilde H_1-H_1)H_0^\dagger - \tilde H_1(SH_0^\dagger +H_0^\dagger T) + \tilde H_1 S(H_0^\dagger-H_0(\eta)) T.
	\end{align*}
	The first term can be controlled as before, since it is like $\|H(\eta)\|_2/\sigma_S(H_0)$. 
}

\textcolor{blue}{I think this is the correct way we should approach the argument. Recall that we would like to bound 
	\[
	\tilde\Psi-\Psi 
	=\tilde H_1\tilde H_0^\dagger - H_1 H_0^\dagger
	=(H_1+H_1(\eta))(H_0+H_0(\eta))^\dagger- H_1 H_0^\dagger. 
	\]
	Let us consider an even simpler situation. If we view $H_0(\eta)$ and $H_1(\eta)$ as small perturbations and pretend they are the same, then above looks similar to the function
	\[
	f(x)=\frac{a+x}{b+x}-\frac{a}{b}.
	\]
	If we do a first order Taylor expansion around $x=0$, we get
	\[
	f(x)
	\approx x\frac{b-a}{b^2}
	\]
	This suggests that 
	\[
	\|\tilde\Psi-\Psi\|
	\leq \frac{\|H_0(\eta)\|_2 \|H_0-H_1\|_2}{\sigma_S(H_0)^2}.
	\]
	So from this informal argument we can conclude the following: it seems that getting a $\sigma_S(H_0)^2$ in the denominator is unavoidable, but what we missed in our previous bound was a term of the form, $\|H_0-H_1\|_2$. If this term is about the same size as the $\sigma_S(H_0)$, then we would get $\|\tilde\Psi-\Psi\|$ bounded above by noise$/\sigma_S(H_0)$, which is equal to or better than the stability of MUSIC.
}

Proposition \ref{propesprit} suggests that it is the best to set $ L=\lfloor (M+1)/2\rfloor$ to balance $\sigma_{\min}(\Phi_L)$ and $\sigma_{\min}(\Phi_{M-L})$. Combining Proposition \ref{propesprit},Lemma \ref{lemmanoise} and Theorem \ref{thm:clump2} gives rise to the following theorem on the stability of ESPRIT in the super-resolution regime.

\begin{theorem}
	\label{thmesprit}
	Let $M$ be an odd integer satisfying $M+1 \ge 2 S^2$, $\alpha>0$ and $\nu>1$.   Suppose $\Omega=\{\omega_j\}_{j=1}^S\subset\T$ consists of $A$ separated clumps $\{\Lambda_a\}_{a=1}^A$, and if $A>1$, assume $\Delta(\Omega)\geq \alpha/(M+1)$ and
	\[
	\min_{m\not=n}\text{dist}(\Lambda_m,\Lambda_n)
	\geq \max_{1\leq a\leq A}\ {20S^{1/2}\lambda_a^{5/2}}{\alpha^{-1/2}(M+1)^{-1}}.
	\]
	For each $1\leq a\leq A$, let $c_a = C_a(\Omega,(M+1)/2)$.
	Suppose $\eta \sim \mathcal{N}(0,\sigma^2 I)$ and  
	$$
	\frac{\sigma}{x_{\min}} <\frac{361(M+1)}{3200 \sqrt{\nu(M+3)\log (M+2)}}   \left(\sum_{a=1}^A C_a^2 \lambda_a^2 \left(\frac{2\lambda_a}{\pi \alpha} \right)^{2\lambda_a-2} \right)^{-1}.
	$$
	Let $\hat\Omega = \{\hat\omega_j\}_{j=1}^S$ be the recovered support by ESPRIT with $L = (M+1)/2$. Then
	$$	
	{\rm dist}_B(\Omega , \hat\Omega) 
	\le
	\left(\frac{20}{19}\right)^4 \frac{160\sqrt{\nu(M+3)\log(M+2)}}{M+1}\cdot
	\frac{x_{\max}}{x_{\min}} 
	\left( \sum_{a=1}^A C_a^2 \lambda_a^2 \left(\frac{2\lambda_a}{\pi \alpha} \right)^{2\lambda_a-2} \right)^2 \cdot \frac{\sigma}{x_{\min}},
	$$
	with probability no less than $1-(M+2)^{-(\nu-1)}$.
\end{theorem}

Theorem \ref{thmesprit} is stated when $M$ is odd for simplicity of the constants, and a similar result holds for even $M$ by setting $L= \lfloor (M+1)/2\rfloor$ in ESPRIT. Theorem \ref{thmesprit} shows that the noise-to-signal ratio that ESPRIT can tolerate at least satisfies
\begin{equation}
\label{espritres1}
\frac{\sigma}{x_{\min}} \propto  \frac{x_{\min}}{x_{\max}} \sqrt{\frac{M}{\log M}}\left(
\sum_{a=1}^A C_a^2 \lambda_a^2 \left(\frac{2\lambda_a}{\pi \alpha} \right)^{2\lambda_a-2}
\right)^{-2}.
\end{equation}
As an example, let us look at the special case where each $\Lambda_a$ contains $\lambda$ equally spaced object with spacing $\alpha/M$ (see Figure \ref{FigDemoClumps1}). In this case, \eqref{espritres1} becomes
\begin{equation}
\frac{\sigma}{x_{\min}} \propto  \frac{x_{\min}}{x_{\max}}  \sqrt{\frac{M}{\log M}} \alpha^{4\lambda-4}
= \frac{x_{\min}}{x_{\max}}  \sqrt{\frac{M}{\log M}} \left(\frac{1}{\rm SRF}\right)^{4\lambda-4},
\label{espritres2}
\end{equation}
which shows that the resolution limit of ESPRIT is exponential in $1/{\rm SRF}$. The exponent $4\lambda - 4$ only depends on the cardinality of clumps instead of the total sparsity $S$. In comparison with our estimate on MUSIC in \eqref{musicres2}, it seems that ESPRIT can tolerate a smaller amount of noise, since the exponent of $\alpha$ is larger in \eqref{espritres1}, and $x_{\min}/x_{\max}$ appears. However, our numerical experiments in Figure \ref{Fig_PhaseTransition} demonstrate that ESPRIT outperform MUSIC in all cases. We conjecture that Proposition \ref{propesprit} is not optimal, and we will consider how to improve it in future work.

}

%\section{Conclusion}

\section*{Acknowledgment}
Wenjing Liao is supported by NSF-DMS-1818751.% and a startup fund from Georgia Institute of Technology.



%\bibliographystyle{IEEEtran}

\bibliographystyle{IEEEtran}
\bibliography{SRlimitFourierbib}

\begin{thebibliography}{10}
\providecommand{\url}[1]{#1}
\csname url@samestyle\endcsname
\providecommand{\newblock}{\relax}
\providecommand{\bibinfo}[2]{#2}
\providecommand{\BIBentrySTDinterwordspacing}{\spaceskip=0pt\relax}
\providecommand{\BIBentryALTinterwordstretchfactor}{4}
\providecommand{\BIBentryALTinterwordspacing}{\spaceskip=\fontdimen2\font plus
\BIBentryALTinterwordstretchfactor\fontdimen3\font minus
  \fontdimen4\font\relax}
\providecommand{\BIBforeignlanguage}[2]{{%
\expandafter\ifx\csname l@#1\endcsname\relax
\typeout{** WARNING: IEEEtran.bst: No hyphenation pattern has been}%
\typeout{** loaded for the language `#1'. Using the pattern for}%
\typeout{** the default language instead.}%
\else
\language=\csname l@#1\endcsname
\fi
#2}}
\providecommand{\BIBdecl}{\relax}
\BIBdecl

\bibitem{donoho1992superresolution}
D.~L. Donoho, ``Superresolution via sparsity constraints,'' \emph{SIAM Journal
  on Mathematical Analysis}, vol.~23, no.~5, pp. 1309--1331, 1992.

\bibitem{demanet2015recoverability}
L.~Demanet and N.~Nguyen, ``The recoverability limit for superresolution via
  sparsity,'' \emph{arXiv preprint arXiv:1502.01385}, 2015.

\bibitem{candes2013super}
E.~J. Cand{\`e}s and C.~Fernandez-Granda, ``Super-resolution from noisy data,''
  \emph{Journal of Fourier Analysis and Applications}, vol.~19, no.~6, pp.
  1229--1254, 2013.

\bibitem{duval2015exact}
V.~Duval and G.~Peyr{\'e}, ``Exact support recovery for sparse spikes
  deconvolution,'' \emph{Foundations of Computational Mathematics}, vol.~15,
  no.~5, pp. 1315--1355, 2015.

\bibitem{li2017elementary}
W.~Li, ``Elementary ${L}^\infty$ error estimates for super-resolution
  de-noising,'' \emph{arXiv preprint arXiv:1702.03021}, 2017.

\bibitem{fannjiang2012coherence}
A.~C. Fannjiang and W.~Liao, ``Coherence-pattern guided compressive sensing
  with unresolved grids,'' \emph{SIAM Journal on Imaging Sciences}, vol.~5,
  no.~1, pp. 179--202, 2012.

\bibitem{liao2016music}
W.~Liao and A.~Fannjiang, ``{MUSIC} for single-snapshot spectral estimation:
  {S}tability and super-resolution,'' \emph{Applied and Computational Harmonic
  Analysis}, vol.~40, no.~1, pp. 33--67, 2016.

\bibitem{moitra2015matrixpencil}
A.~Moitra, ``Super-resolution, extremal functions and the condition number of
  {V}andermonde matrices,'' \emph{Proceedings of the Forty-Seventh Annual ACM
  Symposium on Theory of Computing}, 2015.

\bibitem{morgenshtern2016super}
V.~I. Morgenshtern and E.~J. Candes, ``Super-resolution of positive sources:
  The discrete setup,'' \emph{SIAM Journal on Imaging Sciences}, vol.~9, no.~1,
  pp. 412--444, 2016.

\bibitem{denoyelle2017support}
Q.~Denoyelle, V.~Duval, and G.~Peyr{\'e}, ``Support recovery for sparse
  super-resolution of positive measures,'' \emph{Journal of Fourier Analysis
  and Applications}, vol.~23, no.~5, pp. 1153--1194, 2017.

\bibitem{benedetto2018super}
J.~J. Benedetto and W.~Li, ``Super-resolution by means of {B}eurling minimal
  extrapolation,'' \emph{Applied and Computational Harmonic Analysis}, 2018.

\bibitem{schmidt1986multiple}
R.~Schmidt, ``Multiple emitter location and signal parameter estimation,''
  \emph{IEEE Transactions on Antennas and Propagation}, vol.~34, no.~3, pp.
  276--280, 1986.

\bibitem{li2017stable}
W.~Li and W.~Liao, ``Stable super-resolution limit and smallest singular value
  of restricted fourier matrices,'' \emph{arXiv preprint arXiv:1709.03146},
  2017.

\bibitem{montgomery1974hilbert}
H.~L. Montgomery and R.~C. Vaughan, ``Hilbert's inequality,'' \emph{Journal of
  the London Mathematical Society}, vol.~2, no.~1, pp. 73--82, 1974.

\bibitem{vaaler1985some}
J.~D. Vaaler, ``Some extremal functions in {F}ourier analysis,'' \emph{Bulletin
  of the American Mathematical Society}, vol.~12, no.~2, pp. 183--216, 1985.

\bibitem{batenkov2018stability}
D.~Batenkov, L.~Demanet, G.~Goldman, and Y.~Yomdin, ``Stability of partial
  {F}ourier matrices with clustered nodes,'' \emph{arXiv preprint
  arXiv:1809.00658}, 2018.

\bibitem{wedin1972perturbation}
P.-{\AA}. Wedin, ``Perturbation bounds in connection with singular value
  decomposition,'' \emph{BIT Numerical Mathematics}, vol.~12, no.~1, pp.
  99--111, 1972.

\bibitem{li1998relative}
R.-C. Li, ``Relative perturbation theory: {II}. {E}igenspace and singular
  subspace variations,'' \emph{SIAM Journal on Matrix Analysis and
  Applications}, vol.~20, no.~2, pp. 471--492, 1998.

\bibitem{liao2015multi}
W.~Liao, ``Music for multidimensional spectral estimation: stability and
  super-resolution,'' \emph{IEEE Transactions on Signal Processing}, vol.~63,
  no.~23, pp. 6395--6406, 2015.

\end{thebibliography}


\begin{thebibliography}{1}

\bibitem{IEEEhowto:kopka}
H.~Kopka and P.~W. Daly, \emph{A Guide to \LaTeX}, 3rd~ed.\hskip 1em plus
  0.5em minus 0.4em\relax Harlow, England: Addison-Wesley, 1999.

\end{thebibliography}

\commentout{

}

\end{document}